\documentclass[amsmath,amssymb,twocolumn,showpacs]{revtex4}
\usepackage{amsfonts}
\usepackage{mathrsfs}
\usepackage{graphicx}  % Include figure files

\begin{document}

\title{Symmetry and Explicit Marking of the Critical Phase in Two-Bath Spin-Boson Model}

\author{Yao Yao$^{1,2}$, Nengji Zhou$^{3,4}$, Javier Prior$^5$, and Yang Zhao$^3$\footnote{Electronic address:~\url{YZhao@ntu.edu.sg}}}

\affiliation{
$^1$State Key Laboratory of Surface Physics and Department of Physics, Fudan University, Shanghai 200433, China\\
$^2$Collaborative Innovation Center of Advanced Microstructures, Nanjing University, Nanjing 210093, China\\
$^3$Division of Materials Science, Nanyang Technological University, Singapore 639798, Singapore\\
$^4$Department of Physics, Hangzhou Normal University, Hangzhou 310036, China\\
$^5$Departamento de F\'{i}sica Aplicada, Universidad Polit\'{e}cnica
de Cartagena, Cartagena 30202, Spain }

\date{\today}

\begin{abstract}
The spin-boson model is a paradigm for studying decoherence,
relaxation, entanglement and other effects that arise in a quantum
system coupled to environmental degrees of freedom. At zero
temperature, a localization-delocalization phase transition is known
to exist in the sub-Ohmic regime, where the standard density matrix
renormalization group algorithm is inadequate due to the divergence
in the number of low-frequency modes. This limitation is
circumvented in this work by symmetrically optimizing the phonon
basis and introducing an order parameter accounting for the U(1)
symmetry for a two-bath spin-boson model, by which we are able to
determine the classification and criticality of the phase transition
explicitly. Compared with variational
results, the critical phase is characterized by spontaneous
vanishing of boson displacements in both the baths, resulting in an
accurate phase diagram with three model parameters.
\end{abstract}


\maketitle

%\section{Introduction}

\textit{Introduction- }Much attention has been devoted in recent
years to optical properties of natural photosynthetic systems
\cite{LHC1,LHC2,LHC3,LHC4} and organic photovoltaic devices
\cite{Friend1,bredas,Friend2,Friend3,Guo,ScienceNew}, where the quantum
aspect of excitons and phonons is increasingly recognized to be
essential in boosting the power conversion efficiency. In organic
systems, e.g., delocalization of wave functions is found to be
essential for the dissociation of excitons \cite{Friend1}.
As an excitonic paradigm, a two-level system is
described by a single spin one-half which is coupled to its phonon
environment represented by boson modes. This leads to the celebrated
spin-boson model (SBM) \cite{review1,review2}, providing avenues to
study the phase transition between the localized and delocalized
phases, or from a slightly different perspective, the dynamical
phase transition between the coherent and incoherent phases
\cite{SBM1,SBM2,SBM3,SBM4,SBM5,SBM6,SBM7,Chin3,NRG2,SBM8}.
Despite its simplicity, the SBM is a highly nontrivial model in
nearly all aspects, and currently contention still surrounds the
existence and the precise locations of phase transitions. Designed to
help understand intrinsic mechanisms of coherent exciton
dynamics, every theoretical approach typically works accurately
only in a certain applicable regime, which is far away from the critical point,
preventing the method from addressing issues
related to the phase transition.

The density matrix renormalization group (DMRG) is a powerful
numerical technique to study the low-lying states in strongly
coupled, one-dimensional systems \cite{DMRG}. Similar to that for
the numerical renormalization group \cite{NRG} method, the
orthogonal polynomials theory can be employed to map the SBM to a
one-dimensional chain with only nearest-neighbor coupling
\cite{Chin2}, allowing one to straightforwardly adapt the DMRG
method to study the SBM with only model-free approximations. Over
the last few years, this approach has been extensively used for
detailed studies of the phase transition of SBM
\cite{Chin1,Guo1,Guo2,Guo3,mine1,NJ}. Our recent work was devoted to
the two-bath SBM (TBSBM) to investigate the phase transition in a
comprehensive manner \cite{mine1,NJ}. We have also examined the
dynamics of SBM with the time-dependent DMRG (t-DMRG) algorithm,
compared it with two other established methods, and demonstrated
that a unitary transformation for the state yields reliable,
accurate results \cite{mine2,mine3}.

The approach of the optimal
phonon basis, originally developed to deal with coupled
electron-phonon systems \cite{OB1}, is often adopted to reduce the
dimension of the Fock space of bosons \cite{Guo2,mine1}. Although it has been
utilized for a variety of models \cite{OB2,OB3,OB4}, a
serious problem arises in the context of the TBSBM wherein the
symmetry is numerically broken and the analysis of the phase transition
becomes obstructed \cite{Guo3}. In this work, we circumvent this
problem by symmetrically optimizing the phonon basis and
constructing numerics-friendly operators. Our approach is adapted
specifically for the TBSBM such that the doubly degenerate ground
states can be obtained in a credible manner, and properties of the
phase transition can be studied with sufficient precision.

%\section{Model and Methodology}

%\subsection{Model Hamiltonian}

\textit{Model and methodology-} We consider the TBSBM in which a
single spin is coupled diagonally and off-diagonally to two
independent baths characterized by continuum spectral densities. The
corresponding Hamiltonian can be written as
\begin{equation}
\hat{H}=\sum_{\nu=z,x}\sum_l\left[\omega_l
b^{\dag}_{l,\nu}b_{l,\nu}+\frac{\sigma^{\nu}}{2}\lambda_{l,\nu}(b^{\dag}_{l,\nu}+b_{l,\nu})\right],\label{hami}
\end{equation}
where $\sigma^z$ and $\sigma^x$ are the Pauli operators,
$b^{\dag}_{l,\nu} (b_{l,\nu})$ is the creation (annihilation)
operator of the $l$-th mode of frequency $\omega_l$ in the $\nu$-th
bath ($\nu =z,x$), and $\lambda_{l,\nu}$ represents the
corresponding spin-bath coupling strength.
In the traditional SBM, the bath spectral density has a cut-off
frequency $\omega_c$. For simplicity, the same cut-off frequency is
assigned to the spectral density functions of the two baths in this
work, i.e.,
$J_{\nu}(\omega)=2\pi\alpha_{\nu}\omega^{1-s}_c\omega^{s}
e^{-\omega/\omega_c}$ with $\alpha_{\nu}$ being the dimensionless
coupling strength for the $\nu$ bath and $s$ being the exponent. The
case of $s<1$ corresponds to the sub-Ohmic regime in which the
localized, critical and delocalized phases have been claimed to
exist \cite{Guo2}. We will focus on the sub-Ohmic regime due to its relevance and complexity.

As discussed earlier, the symmetry in the Hamiltonian (\ref{hami})
is of paramount importance to the numerical precision. To
facilitate discussion, we introduce the operators
\begin{eqnarray}
\mathcal{O}^{\pm}_z=\pm\sigma^z{\rm e}^{i\pi\sum_lb^{\dag}_{l,x}b_{l,x}},~~\mathcal{O}^{\pm}_x=\pm\sigma^x{\rm e}^{i\pi\sum_lb^{\dag}_{l,z}b_{l,z}},\label{ozox}
\end{eqnarray}
which commute with the Hamiltonian. Also of interest is their product, i.e.,
\begin{eqnarray}
\mathcal{O}^{\zeta}_y=\mathcal{O}^{\phi}_z\mathcal{O}^{\varphi}_x=i\zeta\sigma^y{\rm e}^{i\pi\sum_l(b^{\dag}_{l,x}b_{l,x}+b^{\dag}_{l,z}b_{l,z})},
\end{eqnarray}
with $\zeta,\phi,\varphi=\pm$ following the common rule of products.
Together with the identity operator $\mathcal{I}^{\pm}$, it can then
be verified straightforwardly that the eight operators
$\mathcal{I}^{\pm},\mathcal{O}^{\pm}_z,\mathcal{O}^{\pm}_x,\mathcal{O}^{\pm}_y$
form a non-abelian group $G$, and its center is represented by
$\{\mathcal{I}^{\pm}\}$. The factor group $G/\{\mathcal{I}^{\pm}\}$
is an abelian group whose irreducible representations are given by
four one-dimensional ones, indicating the U(1) symmetry when
$\alpha_z=\alpha_x$ \cite{Guo3}.
%Specifically speaking, the four representations are characterized by a trivial action of the group center represented by the operators $\mathcal{I}^{\pm}$, as by definition these are two identity operators.
On the other hand, the two-dimensional representation of the
non-abelian group $G$, characterized by a nontrivial central
extension of its factor group, participates in the decomposition in
irreducible representations, resulting in the $\mathbb{Z}_2$
symmetry if $\alpha_z\neq\alpha_x$ \cite{NJ}. Subsequently,
eigenstates of the system and, the ground state in particular, are
doubly degenerate, a novel feature that allows specifications of the
numerical precision in dealing with the symmetry.

%\subsection{DMRG-friendly transformation}

We thus proceed to develop the DMRG algorithm to deal with the
highly symmetrical model. Widely used to study the SBM and related
models \cite{Chin1,Guo1,Guo2,Guo3,mine1,mine2,NJ}, the DMRG approach
starts with the discretization of the boson modes, and employs the
orthogonal polynomials theory to represent the renormalized modes by
a set of boson sites \cite{Chin2}, with a transformed Hamiltonian
\begin{eqnarray}
\tilde{H}&=&\sum_{\nu=z,x}\left[\sqrt{\frac{\eta_{\nu}}{4\pi}}\sigma^{\nu}(b^{\dag}_{0,\nu}+b_{0,\nu})\right.\nonumber\\&+&\left.\sum_i\omega_ib^{\dag}_{i,\nu}b_{i,\nu}+\sum_i(t_ib^{\dag}_{i+1,\nu}b_{i,\nu}+{\rm
h.c.})\right],\label{hami2}
\end{eqnarray}
where $\eta_{\nu}$ is the renormalized coupling calculated from
$\eta_{\nu}=\int_0^{\omega_c}J_{\nu}(\omega)d\omega$. Herein,
$\omega_{i}$ and $t_{i}$ are the frequency and the hopping integral
for the $i$-th site of bosons, respectively, and the expressions for
them can be found in Refs.~\cite{mine1,mine2,NJ}.
%\begin{eqnarray}
%\omega_i=\xi_s(P_i+Q_i),
%\end{eqnarray}
%\begin{eqnarray}
%t_i=-\xi_sP_i(\frac{N_{i+1}}{N_i}),
%\end{eqnarray}
%with
%\begin{eqnarray}
%\xi_s=\frac{(s+1)[1-\Gamma^{-(s+2)}]}{(s+2)[1-\Gamma^{-(s+1)}]}\omega_c,
%\end{eqnarray}
%\begin{eqnarray}
%P_i=\frac{\Gamma^{-i}(1-\Gamma^{-(j+s+1)})^2}{[1-\Gamma^{-(2i+s+1)}][1-\Gamma^{-(2i+s+2)}]},
%\end{eqnarray}
%\begin{eqnarray}
%Q_i=\frac{\Gamma^{-(i+s)}(1-\Gamma^{-j})^2}{[1-\Gamma^{-(2i+s)}][1-\Gamma^{-(2i+s+1)}]},
%\end{eqnarray}
%\begin{eqnarray}
%N^2_i=\frac{\Gamma^{-i(s+1)}(\Gamma^{-1};\Gamma^{-1})^2_i}{[\Gamma^{-(s+1)};\Gamma^{-1}]^2_i[1-\Gamma^{-(2i+s+1)}]},
%\end{eqnarray}
%where
%\begin{eqnarray}
%[a;q]_i=(1-a)(1-aq)\cdot\cdot\cdot(1-aq^{i-1}).
%\end{eqnarray}
%In practise, the two baths are connected with the spin from the left and right hand side, respectively, such that the three comprising portions, namely, the spin plus the two baths, form a one-dimensional chain, which is then DMRG-friendly.

%\subsection{Optimal boson basis}

%\begin{figure}
%\includegraphics[angle=0,scale=0.4]{fig2}
%\caption{Schematic for one step of the DMRG iteration with OB. The left single site is initially with BB and the ground state is calculated for the whole system. Based upon the calculated state, the single site with BB is optimized. Following that the single site moves forward and the above step is redone.}\label{proce}
%\end{figure}

\begin{figure}
\includegraphics[angle=0,scale=0.37]{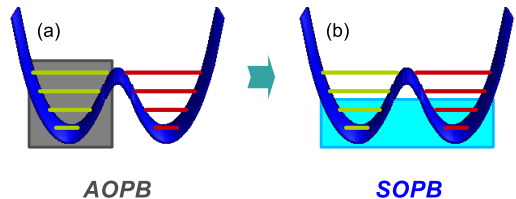}
\caption{Schematic for the two kinds of optimal phonon basis. (a)
The AOPB is shown which takes into account the phonon bases with
shifted displacements \cite{Guo2}. (b) The SOPB is shown which
considers the globally lowest phonon bases.}\label{sch}
\end{figure}

Despite discretization of the spectral densities, the number of the
bare phonon basis in the local Fock space for each renormalized
boson mode is still infinite, hindering numerical calculations. A
remedy is to truncate the Fock space and retain a finite number of
bare phonon states for each mode, resulting in a so-called
restricted phonon basis. Previous tests \cite{mine2} reveal that the
restricted basis method is applicable for the SBM away from the
critical point, but fails to capture the phase-transition properties
in the vicinity of the critical point. An approach employing an
optimal phonon basis (OPB) was adopted by Zhang {\it et al.},
yielding improved results \cite{OB1}. Details of the OPB-adapted
DMRG algorithm can be found in the Supplementary Material. Numerical
difficulties arise, however, in the OPB approach while correctly
tackling the symmetry issues. A seemingly simple solution is to add
an infinitesimal bias and to approach the critical
point asymptotically. However, this would lead to a disastrously large
difference in symmetry between $\alpha_z\neq\alpha_x$ and
$\alpha_z=\alpha_x$ (the critical point) as discussed earlier. To
circumvent this dilemma, we propose two techniques to enforce the
model symmetry.

It is realized that in the implementation of the OPB approach, many
low probability states \ are discarded. As sketched in
Fig.~\ref{sch}(a), e.g., if the sign of the calculated boson
displacement is negative, those states with a positive sign must be
eliminated as they are almost orthogonal to the
negative-displacement states. To recover the symmetry, therefore,
the information in the eliminated states must be fed into the
reduced density matrix, as depicted in Fig.~\ref{sch}(b). In
particular, if $|g_i\rangle$ is the calculated ground state with $i$
as the index of the left free site, we apply the parity operator
$\mathcal{P}_i(\equiv{\rm e}^{i\pi b^{\dag}_{i}b_{i}})$ onto
$|g_i\rangle$, obtaining $|g'_i\rangle$ with an opposite sign of the
boson displacement on $i$-th site. This is numerically feasible as
the $i$-th site is the one that has the Fock space based upon the
bare phonon basis. With the two states obtained, the reduced density
matrix of the $i$-th site is calculated by
\begin{eqnarray}
\rho_i={\rm Tr_E}\left[a|g_i\rangle\langle g_i|+(1-a)|g'_i\rangle\langle g'_i|\right],
\end{eqnarray}
where ${\rm Tr_E}$ denotes the partial trace over all the sites
except the $i$-th one and $a$ is the portion of the state
$|g_i\rangle$ in the reduced density matrix. In the absence of bias,
it is intuitive to set $a$ to $0.5$. We then name the new basis
obtained from the adapted reduced density matrix as the
``symmetrically optimized phonon basis (SOPB)," and accordingly, the
conventional OPB is called the ``asymmetrically optimized phonon
basis (AOPB)." We will show that the optimization procedure based on
the SOPB yields more accurate results.

With one run of the DMRG algorithm, one degenerate ground state is obtained, and
with a second run, we
may obtain another, allowing for a linear combination of the two
\cite{Guo3}. However, symmetry breaking still persists reducing the accuracy and the efficiency.
A sophisticated approach would then be
to apply the operators (\ref{ozox}) to the
calculated ground state. In order to implement this operation,
through the parity operator
$\mathcal{P}\equiv\prod_i\mathcal{P}_i={\rm e}^{i\pi
\sum_ib^{\dag}_{i}b_{i}}$, a unitary transformation on all bosonic
modes must be applied to flip the sign of displacements. As the
bases of the boson modes have been symmetrically optimized, however,
the number operator $\hat{n}_i\equiv b^{\dag}_{i}b_{i}$ ceases to be
diagonal. Moreover, the two degenerate states are almost orthogonal
with each other, so that a simple diagonalization for $\hat{n}_i$
can no longer guarantee numerical precision. In this context, we
introduce a more numerically friendly treatment of the parity
operator by recasting it in the form
\begin{eqnarray}
\mathcal{P}_i=\sum_{n_i={\rm even}}|n_i\rangle\langle n_i|+\sum_{n_i={\rm odd}}{\rm e}^{i\delta\theta}|n_i\rangle\langle n_i|,
\end{eqnarray}
where $|n_i\rangle$ is the eigenstate of the number operator
$\hat{n}_i$ at $i$-th site, and $\delta\theta$ is a small angle.
Herein, following the t-DMRG algorithm \cite{tDMRG}, the angle $\pi$
is divided into many small steps ($\delta\theta$) and the operator is applied
incrementally to the ground state. The operator does not act on the
even-number states, while for the odd-number states one can make
$\mathcal{P}_i(\delta\theta)$ act cumulatively onto the state until
a certain angle $\theta$. If $\theta=\pi$ the action is equivalent
to that of the parity operator, and if $\theta=2\pi$ it is an
identity operator. With this approach, reliable results can be
obtained for all model parameters.

%\section{Results and discussions}

\textit{Results- }In all the calculations we have carried out, the
number of transformed bosonic site is set to be $50$, the number of
bare phonon basis is $16$, and the DMRG truncating number is $64$.
Within these parameters, the error is reduced below $10^{-5}$. The
computation is time-consuming, e.g., on a single 2.13 GHz processor
one run for a set of model parameters needs more than one hundred
hours of CPU time.

%On the other hand, in the previous work of DMRG discussing the phase transition of TBSBM \cite{Guo2}, there are many comparisons of different numerical approaches. Thus in the present work we do not intend to compare our results with that from other methods. We will mainly focus our attention on the precision improvement in the framework of DMRG.

%\subsection{Deep sub-Ohmic regime ($s<0.5$)}

\begin{figure}
\includegraphics[angle=0,scale=1.6]{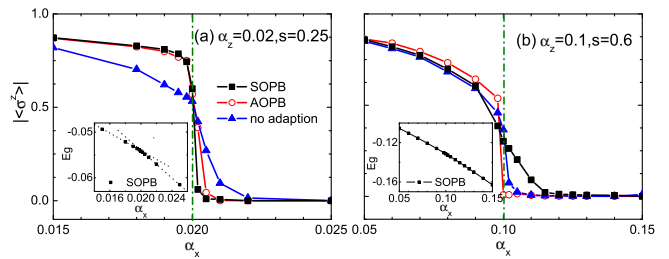}
\caption{$|\langle\sigma^z\rangle|$ versus $\alpha_x$ for three
cases of the basis with (a) $s=0.25,\alpha_z=0.02$ and (b)
$s=0.6,\alpha_z=0.1$. The green dash-dot lines indicate the phase
boundary at $\alpha_x=\alpha_z$. The insets show the ground-state
energy $E_g$ versus $\alpha_x$. The dotted lines in the inset of (a)
are guides for eyes of the linear relationship.}\label{sigmaz}
\end{figure}

%In our previous work \cite{mine1}, we have studied the phase transition of TBSBM in the deep sub-Ohmic regime. A phase transition from delocalized to localized has been determined, and based upon our previous numerical approach a second-order phase transition was addressed. Here, with the newly adapted numerical approach which might give more precise results, we revisit this issue and discuss it in more details.

We first discuss the deep sub-Ohmic regime with $s<0.5$, in which a
transition from localized to delocalized phases has been discussed
in our previous work \cite{mine1}. Fig.~\ref{sigmaz}(a) shows
$|\langle\sigma^z\rangle|$ and the ground-state energy $E_g$ for
various values of $\alpha_x$ with $s=0.25$ and $\alpha_z=0.02$. We
compare three cases, with SOPB, with AOPB, and without any OPB
adaption, and present results in the vicinity of the critical point.
With $|\langle\sigma^z\rangle|$ plotted as a function of the
$\alpha_x$ for the three cases and compared to our previous work
\cite{mine1,NJ}, a much sharper decrease of
$|\langle\sigma^z\rangle|$ from a finite value to zero is found
after the implementation of the parity symmetry. An $\alpha_x$ increment
of $0.0002$ is taken around the critical point, which is numerically equivalent to being
infinitesimal. If the approach with SOPB gives rise to more precise
results, it is implied that the phase transition here is more likely to
be of first order. In addition, the ground-state energy is shown in
the inset for the case with SOPB with the parity symmetry fully
considered. It is observed that there is an obvious kink at the
critical point implying the phase transition. To check the precision
of the SOPB-adapted method, we show in the Supplementary Material
the phase-angle dependence of the bosonic displacement, which serves
as a measure of the precision. The results readily demonstrate the
significant benefits accrued from the precision improvement.

The shallow sub-Ohmic regime with $0.5\leq s<1$ presents even richer
physics. It has been claimed that when $s>0.75$, there is a
so-called critical phase at $\alpha_z=\alpha_x$ with
$\langle\sigma^z\rangle$ and $\langle\sigma^x\rangle$
spontaneously vanishing \cite{Guo2}. From a mean-field analysis \cite{NJ},
however, a similar phenomenon is found for $s > 0.5$ instead of
$s>0.75$. The discrepancy may be attributed to the fact that the mean-field theory
is valid in the weak-coupling limit, while previous DMRG
calculations are applicable in the relatively strong coupling
regime. To resolve the problem, it is necessary to work with a greater
parameter space.

To this end, we first apply the SOPB-adapted approach to the case of
$s=0.6$. In Fig.~\ref{sigmaz}(b), we display
$|\langle\sigma^z\rangle|$ as a function of $\alpha_x$ obtained with
SOPB, with AOPB and without OPB. It is found for the cases with AOPB
and without OPB, $|\langle\sigma^z\rangle|$ shows a rather sudden
change across the critical point. While adopting the SOPB method,
the curve of the $|\langle\sigma^z\rangle|$ across the recognized
critical point becomes much smoother as compared to the other two
cases. This effect implies that the transition is of higher order
than that with the deep sub-Ohmic bath. Moreover, the curve of the
ground-state energy shown in the inset of Fig.~\ref{sigmaz}(b) is
also smooth close to the critical point.

\begin{figure}[bp]
\includegraphics[angle=0,scale=1.6]{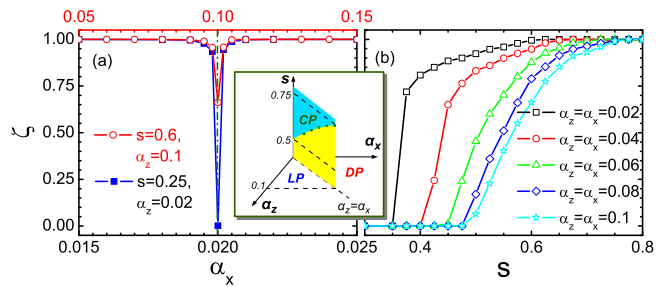}
\caption{(a) The order parameter $\zeta$ versus $\alpha_x$ changing from $0.015$ to $0.035$ for $s=0.25, \alpha_z=0.02$ and changing $0.05$ to $0.15$ for $s=0.6, \alpha_z=0.1$. The dash-dot line denotes the phase boundary. (b) $\zeta$ as a function of $s$ for five sets of $\alpha_z$ and $\alpha_x$. The inset shows the phase diagram consisting of the localized phase (LP), the delocalized phase (DP) and the critical phase (CP).}\label{zeta1}
\end{figure}

In practice, merely considering the magnetization is insufficient
to determine features of the phase transition. Derivation of a more
explicit quantity with the complete information of both
spin and baths is required for this purpose. The argument from the
group theory shows that the operators (\ref{ozox}) are the
generators of the parity symmetry, with eigen-values of $+1$ or
$-1$. Following the group-theory argument, hereafter we calculate
two quantities, $\langle \mathcal{O}_z\rangle$ and
$\langle\mathcal{O}_x\rangle$, involving predictions of both the
spin and the boson components. The calculation becomes possible
because, as tested, the action of the operator $\mathcal{P}$ is
precise based on the SOPB method.

For further clarity, we define an order parameter as
$\zeta=\sqrt{\langle \mathcal{O}_z\rangle^2+\langle
\mathcal{O}_x\rangle^2}$. In the localized and delocalized phases
with $\alpha_z\neq\alpha_x$, due to the orthogonality of the states
$\mathcal{O}_x|g\rangle$ and $\mathcal{O}_z|g\rangle$, either
$\langle \mathcal{O}_z\rangle$ or $\langle\mathcal{O}_x\rangle$
should be 1, such that the quantity $\zeta$ will always be unity.
In the critical phase, as stated, the significance phenomenon is the
spontaneous vanishing of $\langle\sigma^z\rangle$ and
$\langle\sigma^x\rangle$, which obviously results in the vanishing
displacements of bosonic modes. Let $X$ and $Z$ be the displacements
of the bosonic mode in the two baths, respectively, which form a
$X-Z$ plane for all the modes. The critical phase then refers to the
case in which all the modes in the ground state are located at the
origin of the $X-Z$ plane, indicating $\langle\mathcal{O}_z\rangle$,
$\langle\mathcal{O}_x\rangle$ and $\zeta$ are all unity. Out of the
critical phase, however, the continuous U(1) symmetry dominates the
critical point with $\alpha_z=\alpha_x$. It implies that the bosonic
modes have certain displacements and deviate from the origin,
leading to $\zeta$ smaller than one. The order parameter $\zeta$
accounting for the U(1) symmetry is thus a measurement of the
effective distance of this deviation.

\begin{figure}
\includegraphics[angle=0,scale=1.6]{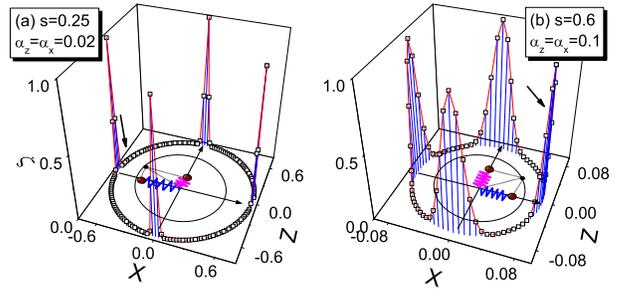}
\caption{The order parameter $\zeta$ calculated by the variational approach with rotational optimization shown in the $X-Z$ plane for (a) $s=0.25, \alpha_z=\alpha_x=0.02$ and (b) $s=0.6, \alpha_z=\alpha_x=0.1$. The radius of the circle in the $X-Z$ plane is determined by averaged boson displacements. The schematics in the $X-Z$ plane shows the displacements with respect to the calculated ground states, which are pointed out by the arrows as well.}\label{Theta}
\end{figure}

Fig.~\ref{zeta1}(a) shows the $\alpha_x$ dependence of $\zeta$ with
$\alpha_z=0.02$ and 0.1 for $s=0.25$ and $0.6$, respectively. It is
clear that for the deep sub-Ohmic case, $\zeta$ undergoes a sudden
change between 1 and 0 close to the critical point
($\alpha_x=\alpha_z$). In terms of the above arguments, this finding
proves the absence of a critical phase and clarifies the first-order
phase transition for the sub-Ohmic bath. More interesting is the
shallow sub-Ohmic case, in which $\zeta$ is between 0 and 1 at the
critical point implying the possible appearance of the critical
phase. In order to see the transition point between the localized
and critical phases, we show in Fig.~\ref{zeta1}(b) the
$s$-dependent $\zeta$ for five $\alpha_z$'s right at
$\alpha_x=\alpha_z$. It is found that $\zeta$ vanishes if $s$ is small,
while it becomes unity if $s$ is larger than a certain value.
More importantly, the larger the $\alpha_z$, the
larger the value of $s$ for which $\zeta$ equals to one. In
particular, for $\alpha_z=0.1$, the transition point $\zeta$ becomes
unity is located at $s=0.75$ which is in agreement with that of
the previous work \cite{Guo2}. On the other hand, a scaling analysis
of $\alpha_z$ to the weak-coupling limit shows the transition point
of $\zeta=1$ moving to $s=0.5$ consistent with the expectation of
the mean-field analysis \cite{NJ}. Based on the transition points of
$\zeta$, we draw a phase diagram shown in the inset of
Fig.~\ref{zeta1}, where the boundaries of the localized,
delocalized and critical phases are explicitly determined.

Finally, to ensure that the vanishing $\zeta$ is not caused by
numerical instability, we have also carried out the calculations
based on variational theory \cite{NJ}. Via this method, we first
produce a state with the lowest energy, which is not necessarily the
ground state especially at the critical point with high numerical
noise. Afterward, we rotate the boson states with a phase angle
$\Theta$ following the same procedure as that adopted for the DMRG
calculations. This approach is justified because at the critical
point the system obeys U(1) symmetry. By comparing the energies of
the states, a real ground state can be found subsequently. Details
of this approach are provided in the Supplementary Material. For
each state with $\Theta$, we calculate the average displacements $X$
and $Z$ of both baths and the value of $\zeta$ accordingly, as shown
in Fig.~\ref{Theta}. To a large extent, $\zeta$ is also found to vanish by the variational
approach, lending support to our DMRG
results. More importantly, in the deep sub-Ohmic regime, $\zeta$
completely vanishes for a majority of the cases, implying the states
to be almost orthogonal to each other. The location of the real
ground state has been indicated by the arrows in Fig.~\ref{Theta},
in agreement with the DMRG results. Moreover, we have also compared
the average displacement of all the bosonic modes with the DMRG
results and the results are presented in Supplementary Material. The
displacement in the case $s=0.6$ is much smaller than that of
$s=0.25$, implying the ground state of $s=0.6$ to be closer to the
origin of the $X-Z$ plane rather than that of $s=0.25$. This finding
elucidates why $\zeta$ is more likely to be nonzero for $s=0.6$ than
for $s=0.25$.

%\section{Conclusion}

\textit{Conclusion- }In summary, we have employed the DMRG algorithm
to study the TBSBM with both deep and shallow sub-Ohmic bosonic
spectral densities. The numerical approach is adapted with the SOPB, and the
parity operator is optimally constructed. It is
found that through this adaptation, the numerical precision is greatly improved in both sub-Ohmic regimes.
Using the SOPB method, we investigate the spin population and the ground-state energy. The results show that for
$s=0.25$ both the two quantities change significantly, supporting a
feature of the first-order phase transition, while for $s=0.6$ the
changes become smoother indicating a higher order phase transition.
We have also put forth a newly defined order parameter $\zeta$, based on which
various features of the critical phase are
discussed and the phase diagram is explicitly obtained.
It is concluded that the SOPB-adapted DMRG algorithm
is well-suited to handle the complexity of the phase transitions in the SBM.
This robust approach is expected to be extended to tackle
other issues, such as the real-time dynamics of the SBM.

{\it acknowledgments-}
The authors gratefully acknowledge support from the NSF of China
(Grant Nos.~91333202, 11134002 and 11104035), the National Basic
Research Program of China (Grant No.~2012CB921401), and the
Singapore National Research Foundation through the Competitive
Research Programme (CRP) under Project No.~NRF-CRP5-2009-04.


\begin{thebibliography}{99}

\bibitem{LHC1} G. S. Engel, T. R. Calhoun, E. L. Read, T. K. Ahn, T. Mancal, Y.
C. Cheng, R. E. Blankenship, and G. R. Fleming, Nature (London)
\textbf{446}, 782 (2007).

\bibitem{LHC2} I. P. Mercer, Y. C. El-Taha, N. Kajumba, J. P. Marangos, J.W. G.
Tisch, M. Gabrielsen, R. J. Cogdell, E. Springate, and E. Turcu,
Phys. Rev. Lett. \textbf{102}, 057402 (2009).

\bibitem{LHC3} R. E. Fenna and B. W. Matthews, Nature (London) \textbf{258}, 573 (1975).

\bibitem{LHC4} G. Panitchayangkoon, D. Hayes, K. A. Fransted, J. R. Caram, E.
Harel, J. Wen, R. E. Blankenship, and G. S. Engel, Proc. Natl. Acad.
Sci. U.S.A. \textbf{107}, 12766 (2010).

\bibitem{Friend1} A. A. Bakulin, A. Rao, V. G. Pavelyev, P. H. M.
van Loosdrecht, M. S. Pshenichnikov, D. Niedzialek, J. Cornil, D.
Beljonne, and R. H. Friend, Science \textbf{335}, 340 (2012).

\bibitem{bredas} J. -L. Bredas, Science \textbf{343}, 492 (2014).

\bibitem{Friend2} A. Rao, P. C. Y. Chow, S. G\'{e}linas, C. W. Schlenker,
C. Z. Li, H. L. Yip, A. K. -Y. Jen, D. S. Ginger, and R. H. Friend,
Nature (London) \textbf{500}, 435 (2013).

\bibitem{Friend3} S. G\'{e}linas, A. Rao, A. Kumar, S. L. Smith, A. W. Chin,
J. Clark, T. S. van der Poll, G. C. Bazan, and R. H. Friend, Science
\textbf{343}, 512 (2014).

\bibitem{Guo} J. Guo, H. Ohkita, H. Benten, and S. Ito, J. Am. Chem. Soc. \textbf{132}, 6154 (2010).

\bibitem{ScienceNew} S. M. Falke, C. A. Rozzi, D. Brida, M. Maiuri, M. Amato, E. Sommer, A. De Sio,
A. Rubio, G. Cerullo, E. Molinari, and C. Lienau, Science
\textbf{344}, 1001 (2014).


\bibitem{review1} A. J. Leggett, S. Chakravarty, A. T. Dorsey, P. A. Fisher, A.
Garg, Rev. Mod. Phys. \textbf{59}, 1 (1987).

\bibitem{review2} U. Weiss, \textit{Quantum Dissipative Systems}, 3rd ed.
(World Scientific, Singapore, 2007).

\bibitem{SBM1} D. Kast and J. Ankerhold, Phys. Rev. Lett. \textbf{110}, 010402 (2013).

\bibitem{SBM2} R. Egger and C. H. Mak, Phys. Rev. B \textbf{50}, 15210
(1994); L. Muhlbacher, J. Ankerhold, J. Chem. Phys. \textbf{122},
184715 (2005); L. Muhlbacher, J. Ankerhold, A. Komnik, Phys. Rev.
Lett. \textbf{95}, 220404 (2005).

\bibitem{SBM3} H. Wang and M. Thoss, New J. Phys. \textbf{10}, 115005
(2008); Chem. Phys. \textbf{370}, 78 (2010).


\bibitem{SBM4} Z. L\"{u} and H. Zheng, J. Chem. Phys. \textbf{136}, 121103
(2012).

\bibitem{SBM5} A. Ishizaki and G. R. Fleming, J. Chem. Phys. \textbf{130}, 234111
(2009).

\bibitem{SBM6} H. Hossein-Nejad and G. D Scholes, New J. Phys. \textbf{12}, 065045 (2010).

\bibitem{SBM7} D. P. S. McCutcheon, N. S. Dattani, E. M. Gauger, B. W. Lovett, and A. Nazir, Phys. Rev. B \textbf{84}, 119903 (2011).

\bibitem{Chin3} A. W. Chin, J. Prior, S. F. Huelga, and M. B. Plenio
Phys. Rev. Lett. \textbf{107}, 160601 (2011).

\bibitem{NRG2} Y. Y. Zhang, Q. H. Chen, and K. L. Wang, Phys. Rev. B
\textbf{81}, 121105(R) (2010).

\bibitem{SBM8} P. Nalbach and M. Thorwart, Phys. Rev. B \textbf{81}, 054308
(2010).


\bibitem{DMRG} S. R. White, Phys. Rev. B 48, 10345 (1993).

\bibitem{NRG} R. Bulla, H. J. Lee, N. H. Tong, and M. Vojta, Phys.
Rev. B \textbf{71}, 045122 (2005).

\bibitem{Chin2} A. W. Chin, \'{A}. Rivas, S. F. Huelga, and M. B. Plenio,
J. Math. Phys. \textbf{51}, 092109 (2010).

\bibitem{Chin1} J. Prior, A. W. Chin, S. F. Huelga, and M. B. Plenio, Phys. Rev. Lett. \textbf{105}, 050404
(2010).

\bibitem{Guo1} C. Guo, A. Weichselbaum, S. Kehrein, T. Xiang, and J. von
Delft, Phys. Rev. B \textbf{79}, 115137 (2009).

\bibitem{Guo2} C. Guo, A. Weichselbaum, J. von Delft, and M. Vojta, Phys. Rev. Lett. \textbf{108}, 160401 (2012).

\bibitem{Guo3} B. Bruognolo, A. Weichselbaum, C. Guo, J. von Delft, I. Schneider, and M. Vojta, arXiv:1410.3821 (unpublished).

\bibitem{mine1} Y. Zhao, Y. Yao, V. Chernyak, and Y. Zhao, J. Chem. Phys. \textbf{140}, 161105 (2014).

\bibitem{NJ} N. Zhou, L. Chen, Y. Zhao, D. Mozyrsky, V. Chernyak, and Y. Zhao, Phys. Rev. B \textbf{90}, 155135 (2014).

\bibitem{mine2} Y. Yao, L. Duan, Z. L\"{u}, C. Q. Wu, and Y. Zhao, Phys. Rev. E \textbf{88}, 023303 (2013).

\bibitem{mine3} Y. Yao, Phys. Rev. B \textbf{91}, 045421 (2015).

\bibitem{OB1} C. Zhang, E. Jeckelmann, and S. R. White, Phys. Rev. Lett. \textbf{80}, 2661 (1998).

\bibitem{OB2} B. Friedman, Phys. Rev. B \textbf{61}, 6701 (2000).

\bibitem{OB3} A. Wei$\beta$e, H. Fehske, G. Wellein, and A. R. Bishop, Phys. Rev. B \textbf{62}, R747 (2000).

\bibitem{OB4} W. Q. Ning, H. Zhao, C. Q. Wu, and H. Q. Lin, Phys. Rev. Lett. \textbf{96}, 156402 (2006).

\bibitem{tDMRG} A. J. Daley, C. Kollath, U. Schollw\"{o}ck, and G. Vidal, J. Stat.
Mech.: Theor. Exp. (2004) P04005; S. R. White and A. E. Feiguin,
Phys. Rev. Lett. \textbf{93}, 076401 (2004).


\end{thebibliography}
\end{document}

\title{Supplementary Materials: Symmetry and Explicit Marking of the Critical Phase in Two-Bath Spin-Boson Model}

\author{Yao Yao$^{1,2}$, Nengji Zhou$^{3,4}$, Javier Prior$^5$, and Yang Zhao$^3$\footnote{Electronic address:~\url{YZhao@ntu.edu.sg}}}

\affiliation{
$^1$State Key Laboratory of Surface Physics and Department of Physics, Fudan University, Shanghai 200433, China\\
$^2$Collaborative Innovation Center of Advanced Microstructures, Nanjing University, Nanjing 210093, China\\
$^3$Division of Materials Science, Nanyang Technological University, 50 Nanyang Avenue, Singapore 639798\\
$^4$Department of Physics, Hangzhou Normal University, Hangzhou 310036, China \\
$^5$ Departamento de F{\'i}sica Aplicada, Universidad Polit{\'e}cnica de Cartagena, Cartagena 30202, Spain
}

\date{\today}

\maketitle

\section{Optimal phonon basis}

\begin{figure}
\includegraphics[angle=0,scale=0.4]{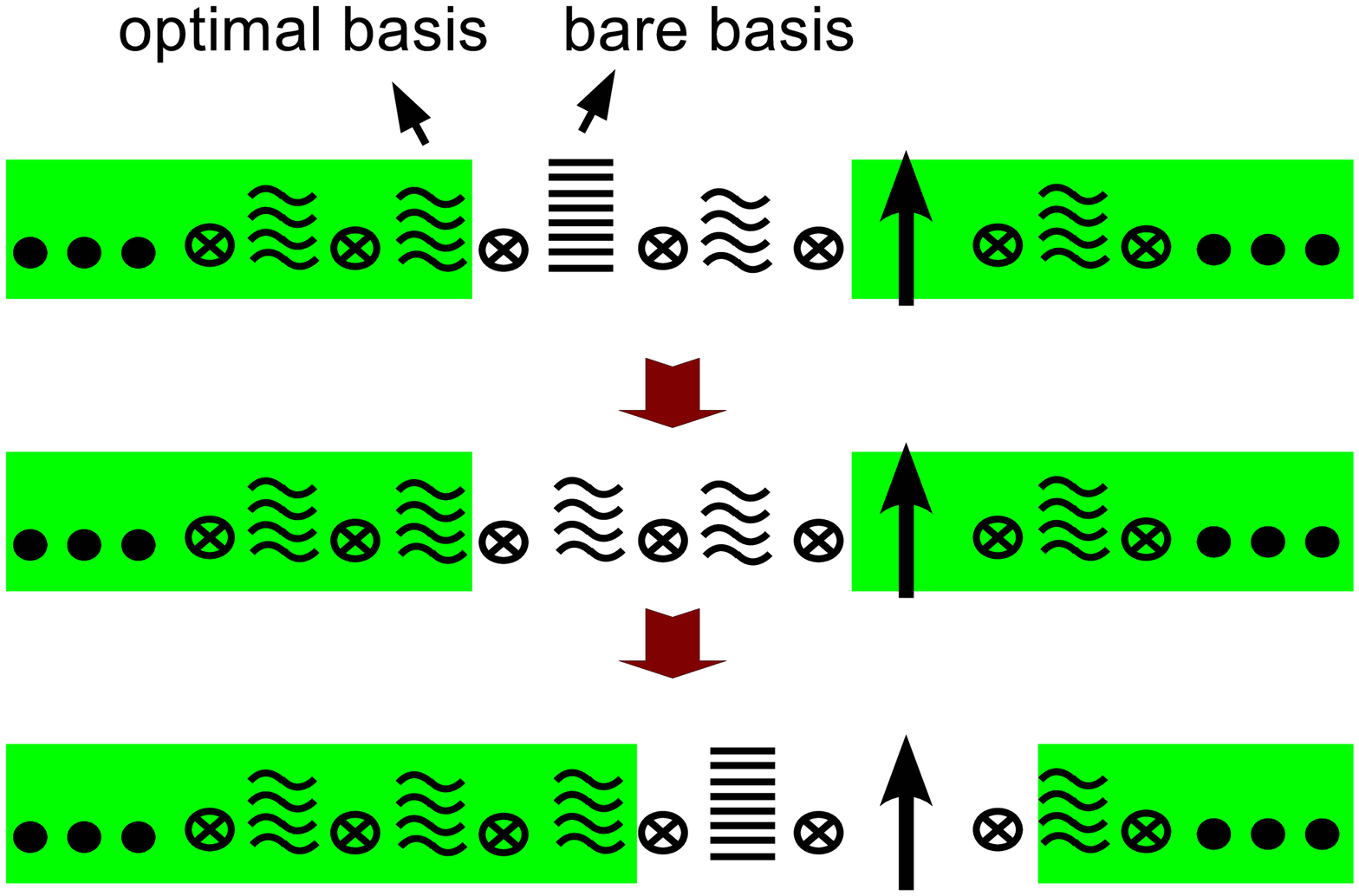}
\caption{Schematic for one step of the DMRG iteration with OPB. The left single site is initially with bare basis and the ground state is calculated for the whole system. Based upon the calculated state, the single site with bare basis is optimized. Following that the single site moves forward and the above step is redone.
The large vertical arrow in the chain represents a spin.}\label{proce}
\end{figure}

The calculation procedure of the OPB-adapted DMRG algorithm, as sketched in Fig.~\ref{proce}, can be briefly summed up as follows. Firstly, we apply the usual DMRG procedure with restricted basis to obtain the state with the lowest energy which may be a rough estimate of the true ground state energy of the system \cite{DMRG}. To accelerate the computations, it is useful to add a very small bias $\epsilon$ to the system in this step to lift the degeneracy. The sign of the bias is not important in this step, because in the steps later on both degenerate states will be obtained and the influence of the small bias will be eliminated by the scaling analysis. Secondly, the DMRG iteration is continued, and the basis of the Fock space on the left single site is replaced by the bare phonon one during each step of the iteration. The number of bare basis ($N_B$) is much larger than that of the restricted basis ($N_R$), ensuring the convergency of the local energy. Thirdly, a new state with a much lowered energy with respect to the bare basis is obtained. Based upon this state, we calculate the reduced density matrix of the left single site and follow the idea of DMRG to discard those bases with low probability. The new reduced basis for the local Fock space is the OPB, carrying as much information as the bare phonon basis. After one iteration, all the boson sites are optimized with OPB, and then the system energy is minimized using the usual DMRG procedure. Finally, a state with an energy very close to that of the real ground state is obtained, with which we can calculate all desired observables.

\section{Effective displacement of bosons}

\subsection{Deep sub-Ohmic regime ($s\leq0.5$)}

In order to examine the reliability of our numerical approach, we apply the operator $\mathcal{P}_i(\delta\theta)$ repeatedly onto the ground state $|g\rangle$ to obtain a new state $|\theta\rangle$. As discussed, the states with opposite displacements are orthogonal to each other, such that the new state with $\theta=\pi$ is orthogonal to the ground state, i.e., $\langle g|\theta = \pi \rangle=0$. Therefore, if the information of the state with the opposite displacement is not taken into account in the procedure of the numerical optimization, the action of the operator $\mathcal{P}_i(\delta\theta)$ will give rise to results completely out of line with expectations. The examination of this operator is a critical criterion to determine whether our numerical approach is able to attain the double degeneracy, and thus the phase transition.

\begin{figure}
\includegraphics[angle=0,scale=1.65]{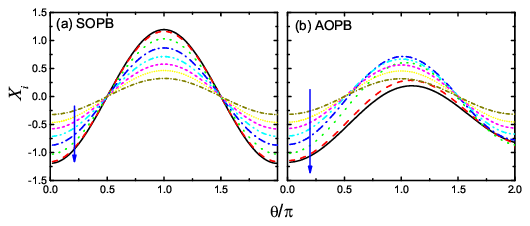}
\caption{The effective displacement of boson on the odd sites that are closest to the spin versus the rotation angle with (a) the SOPB and (b) the AOPB. The parameters are $s=0.25, \alpha_z=\alpha_x=0.02$, and $N_B=16$. Different colours correspond to different sites, and the arrows indicate the orientation away from the spin.}\label{deep2}
\end{figure}

Based upon the state $|\theta\rangle$, we calculate the displacements of every boson mode $X_i({\theta})(\equiv 1/\sqrt{2}\langle b_i^{\dag}+b_i\rangle$) with both the symmetrically optimized phonon basis (SOPB) and asymmetrically optimized phonon basis (AOPB), and results shown in Fig.~\ref{deep2} are for the case of $\alpha_x=0.02$ and $8$ odd sites that are closest to the spin. It is found that with SOPB, displacements all switch signs when $\theta=\pi$, while the same signs are recovered when $\theta=2\pi$. On the other hand, the AOPB displacements are found to be smaller than those of SOPB. More importantly, after a $2\pi$ rotation the displacements fail to return to their original positions, implying anomalies in the curves when $\theta$ becomes large. We also calculate the energy (not shown), which is directly related to the displacement of the boson modes, and for the two cases we also observe
that the energy of SOPB case at $\theta=2\pi$ reproduces the original value while that of AOPB case does not. These results clearly show great improvements when the parity symmetry is explicitly taken into account. In addition, the deviation $|\Delta X_i|$ for the displacement of $\theta=2\pi$ from that of $\theta=0$ is a perfect measure of the numerical precision. The value of $|\Delta X_i|$ is found to be smaller than $10^{-5}$ in Fig.~\ref{deep2}, indicating the results in the case of $s=0.25$ and $\alpha_x=0.02$ are reliable. We have also checked the value of $\langle g|\theta=\pi\rangle$ which is of the same order as the deviation.

\subsection{Shallow sub-Ohmic regime ($0.5 < s <1$)}

\begin{figure}
\includegraphics[angle=0,scale=1.65]{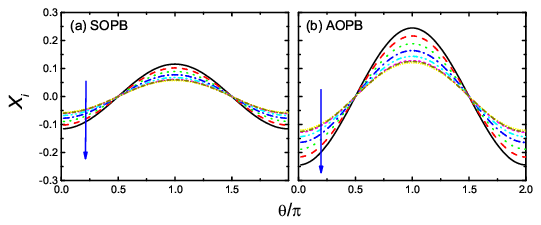}
\caption{The effective displacement of boson on the odd sites that are closest to the spin versus the rotation angle with (a) the SOPB and (b) the AOPB. The parameters are $s=0.6, \alpha_z=\alpha_x=0.1$, and $N_B=16$. Different colours correspond to different sites, and the arrows indicate the orientation away from the spin.}\label{sha2}
\end{figure}

In order to indicate the benefits of the SOPB method in the shallow sub-Ohmic regime, we show in Fig.~\ref{sha2} the phase angle dependence of the displacements for both SOPB and AOPB cases similar to that in the previous subsection. At variance to the deep sub-Ohmic case, the shallow sub-Ohmic bath produces no anomalies for the two bases, as the displacements show perfect rotations in line with the phase angle implying great precision. On the other hand, the AOPB method gives rise to much larger displacements than those from the SOPB method. Similar with the analysis in the deep sub-Ohmic case, a large displacement here would result in a large polarization of the spin, and the AOPB spin population shows a sudden change at the transition point,  while the SOPB one shows a continuous crossover. The latter case is more accurate according to our previous mean-field analysis. Therefore, the applicable extent of the method is determined, that is, the SOPB-adapted method works well for both deep and shallow sub-Ohmic cases while dealing with the phase transition.

\section{The variational approach for the TBSBM}

\subsection{The rotational optimization}

In order to make a comparison, we adopt the variational approach \cite{NJ} to calculate the quantities. The variational approach is based on the trial wavefunction
\begin{eqnarray}
|\psi \rangle & = & |+\rangle \sum_{n}^N A_n \exp\left[ \sum_{l}^{2M}\left(f_{n,l}b_l^{\dag} - {\rm H.c.}\right)\right] |0\rangle_{\textrm{ph}} \nonumber \\
              & + & |-\rangle \sum_{n}^N B_n \exp\left[ \sum_{l}^{2M}\left(g_{n,l}b_l^{\dag} - {\rm H.c.}\right)\right] |0\rangle_{\textrm{ph}},
\label{vmwave}
\end{eqnarray}
where $|+\rangle$ ($|-\rangle$) stands for the spin up (down) state in the $Z$-direction, $\rm H.c.$ denotes the Hermitian conjugate, $|0\rangle_{\textrm{ph}}$ is the vacuum state of the phonon bath, $M$ and $N$ represent the numbers of the bath modes and coherent superposition states, respectively, and $A_n, B_n, f_{n,l}$ and $g_{n,l}$ are the variational parameters. Herein, the variables $n$ and $l$ represent the rank of the coherent superposition state and the label of the bosonic mode, respectively. By defining the norm of the wave function as $D=\langle\psi|\psi\rangle$ and the energy $E$ as $E=H/D$ with $H\equiv\langle\psi|\hat{H}|\psi\rangle$, one can derive the self-consistency equations as the variational conditions, namely
\begin{equation}
\frac{\partial H}{\partial \{\xi_{i}\}} - E\frac{\partial D}{\partial \{\xi_{i}\}} = 0,
\label{vmit}
\end{equation}
where $\xi_i$ denotes one of the variational parameters. The ground state is then searched by using this variational iteration technology.

Due to the numerical errors, however, the state $|\psi\rangle$ found by the variational approach may be a meta-stable state. This situation is quite similar to that in the DMRG calculation. In particular, since the coherent states are used as the trial wave function, the calculated critical point will always deviate from the theoretical point $\alpha_x=\alpha_z$. Motivated by the SOPB-adapted DMRG algorithm, we then incorporate the symmetry into the variational approach. We construct a rotational operator $\hat{T}(\Theta)=\exp(i\Theta \hat{S})$, where
\begin{equation}
\hat{S}=\frac{1}{2}\sigma_y + i\sum_{l}\left(b_{l,x}b_{l,z}^{\dag}-b_{l,x}^{\dag}b_{l,z}\right)
\label{generator}
\end{equation}
is the generator of the U($1$) symmetry. It is worth noting that, there is a hopping term between the two baths in the generator $\hat{S}$, making it hard to implement the U($1$) symmetry in the DMRG algorithm which will be discussed later on. By applying the operator $\hat{T}(\Theta)$ onto the state $|\psi\rangle$, we are able to find the ground state with the lowest energy in the $X-Z$ plane, which is much more reliable than that found by the variational method without the optimization.

\begin{figure}[tbp]
\includegraphics[angle=0,scale=0.7]{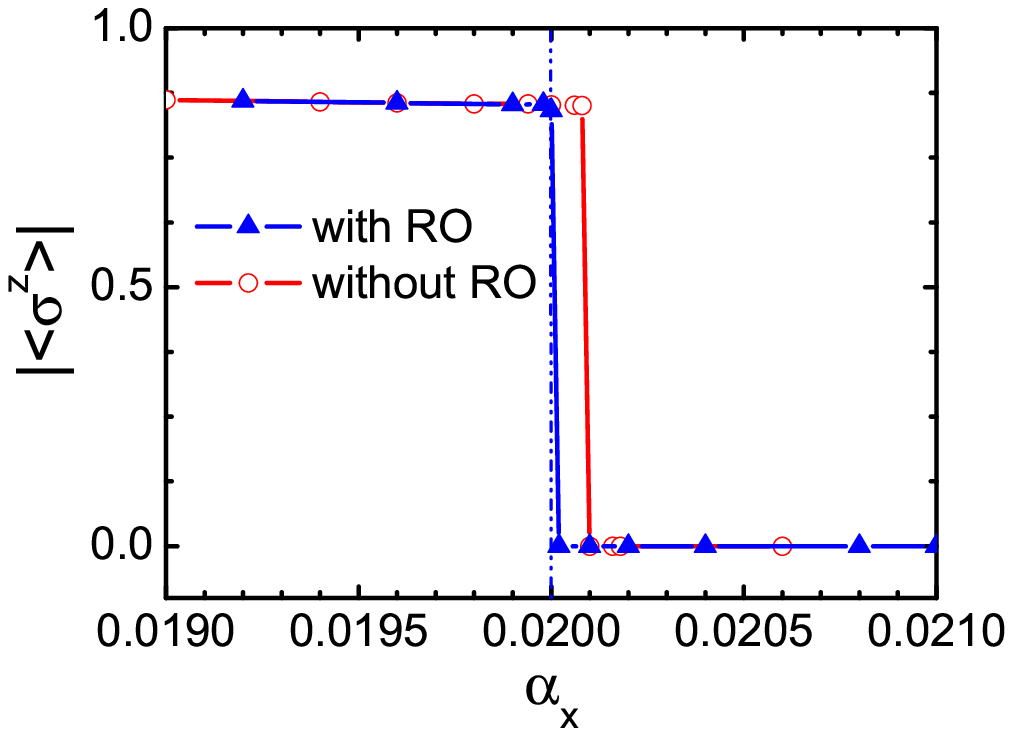}
\caption{$|\langle \sigma_z\rangle|$ calculated by the variational approach as a function of $\alpha_x$ ranging from $0.018$ to $0.021$ for $s=0.25$ and $\alpha_z=0.02$ with and without the rotational optimization (RO). The blue dash-dotted line indicates the critical point at $\alpha_{x}=\alpha_z$. }
\label{sigZV}
\end{figure}

\begin{figure}[tbp]
\includegraphics[angle=0,scale=0.7]{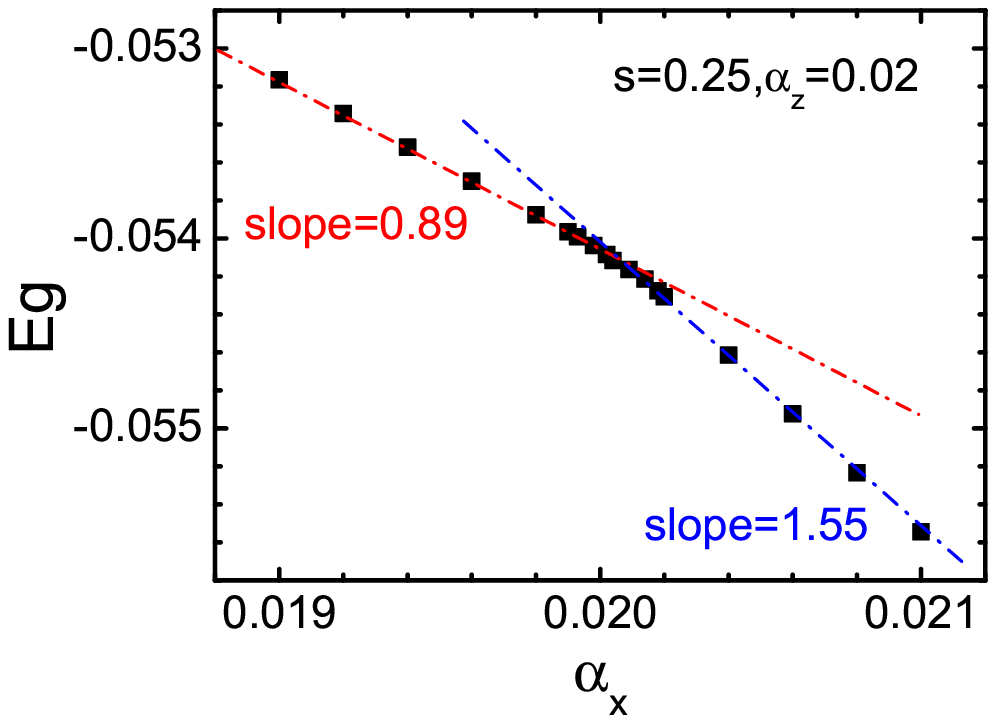}
\caption{The ground state energy $E_g$ as a function of $\alpha_x$ at $s=0.25$ and $\alpha_z=0.02$ calculated by the variational approach with rotational optimization. The dashed lines represent the linear fits in both the localized (red) and delocalized phases (blue).}
\label{enV}
\end{figure}

In the case with $\alpha_z=0.02$ and $s=0.25$, a sharp jump of the magnetization $|\langle \sigma_z\rangle|$ has been found at the critical point $\alpha_{x,c} \approx 0.02$, similar to the results obtained by the DMRG algorithm with SOPB. To locate the critical point explicitly, the behavior of the magnetization
in a much narrower range $[0.019, 0.021]$ is investigated. As shown in Fig.~\ref{sigZV}, the value of the critical point $\alpha_{x,c}=0.02000(1)$ is located with the rotational optimization, an improvement over $a_{x,c}=0.02008(1)$ obtained without the optimization, since the theoretical value of $\alpha_{x,c}=\alpha_{z}=0.02$ is expected.
It indicates that the rotational optimization makes the variational approach more precise in determining the critical point. Also calculated is the ground state energy $E_g$, as shown in Fig.~\ref{enV}. The slope of the curve $E_g(\alpha_x)$ is comparable with that obtained by DMRG, though the explicit value of $E_g$ is slightly larger than that from DMRG, e.g., $E_g=-0.12917$ for the variational method and $E_g=-0.12923$ for the DMRG algorithm at $s=0.6$ and $\alpha_x=\alpha_z=0.1$, implying the high precision of the DMRG algorithm. Moreover, two distinct values of the slopes, $0.89$ and $1.55$, are obtained for the localized and delocalized phases, respectively, indicating a first order phase transition.

\subsection{Order parameter calculated by the variational approach with rotational optimization}

\begin{figure}[bp]
\includegraphics[angle=0,scale=0.7]{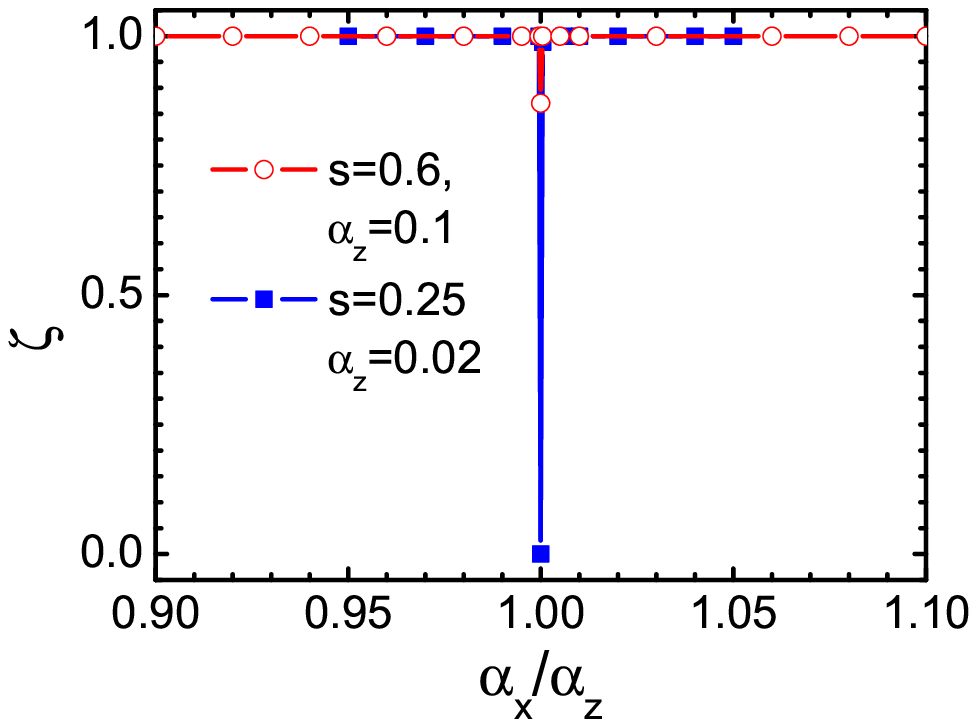}
\caption{The order parameter $\zeta$ calculated by the variational approach with the rotational optimization as a function of the ratio $\alpha_x/\alpha_z$ for the two cases of $s=0.25, \alpha_z=0.02$ and $s=0.6, \alpha_z=0.1$, respectively.}
\label{zeta}
\end{figure}

The order parameter $\zeta$ is also calculated by the variational approach with the rotational optimization, in comparison with that from DMRG algorithm with SOPB. As shown in Fig.~\ref{zeta}, $\zeta$ is plotted for the two cases of $s=0.25, \alpha_z=0.02$ and $s=0.6, \alpha_z=0.1$. In the deep sub-Ohmic regime with $s=0.25<0.5$, $\zeta$ is found to be vanishing at the critical point $\alpha_{x}/\alpha_z=1.0$, i.e, $\zeta(1.0)=0$. On the other hand, a slight decrease from $\zeta=1$ is found at the critical point in the shallow sub-Ohmic regime with $s=0.6$. The result is in good agreement with that obtained by DMRG, indicating the potency of the SOPB-adapted DMRG algorithm.

\begin{figure}[tbp]
\includegraphics[angle=0,scale=0.7]{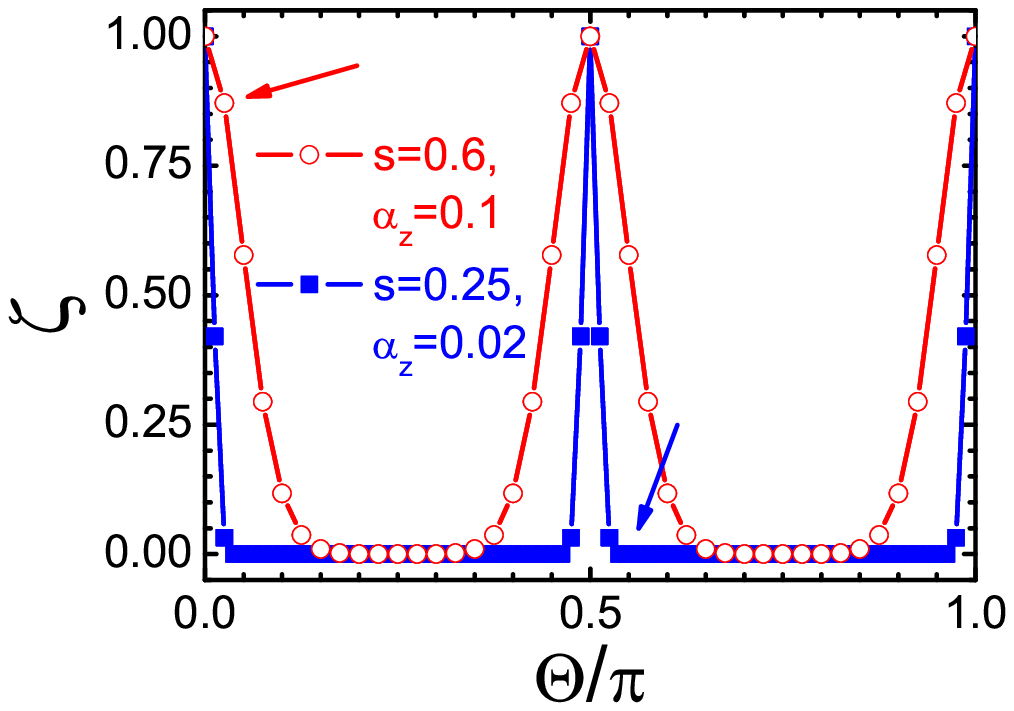}
\caption{The order parameter $\zeta$ calculated by the variational approach with the rotational optimization is displayed as a function of $\Theta/\pi$ for the two cases of $s=0.25, \alpha_x=\alpha_z=0.02$ and $s=0.6, \alpha_x=\alpha_z=0.1$ in the deep and shallow sub-Ohmic regimes, respectively.
The arrows indicate the locations of the calculated ground states.}
\label{circle}
\end{figure}

To further understand the order parameter $\zeta$ at the critical point, the behavior of $\zeta(\Theta)$ is investigated on the states $\hat{T}(\Theta)|\psi\rangle$ with the rotational optimization, as shown in Fig.~\ref{circle}. For $s=0.25$ in the localized phase, sharp peaks of $\zeta(\theta)$ are found at $\Theta/\pi \approx n/2$ ($n=0, 1,2$) with a small peak width $\Delta \Theta_1$ defined as the size of the parity-symmetry regime with $\zeta>0$. It suggests that the ground state is localized in a corner of the $ X - Z$ plane, where $X$ and $Z$ corresponding to the diagonal and off-diagonal coupling baths, respectively. For $s=0.6 > 0.5$ in the critical phases, the width of the peak $\Delta \Theta_2$ increases visibly. Interestingly, the ratio of these two peak widths $\Delta\Theta_2/\Delta\Theta_1=6.0$ is consistent with the ratio of ground state central angles being $6.3$ in the $ X - Z$ plane, indicating that the width of the peak is determined by the central angle. Since the ground state marked by the arrow is located nearby the peak, the value of $\zeta$ in the ground state increases with the width of the peak. It elucidates why $\zeta$ is more likely to be nonzero for $s=0.6$ than that for $s=0.25$ in Fig.~\ref{zeta}.

\section{U(1) symmetry of TBSBM with Single mode}

At $\alpha_x=\alpha_z$, the entire system obeys the U($1$) symmetry with the generator expressed in Eq.~(\ref{generator}). Very recently, Bruognolo {\it et al}$.$~tried to incorporate the U(1) symmetry explicitly into the DMRG algorithm \cite{Guo3}. It is not easy done since there is a hopping term between the two bosonic baths in the generator. To implement the symmetry, they make a transformation in the Hamiltonian of TBSBM to mix the boson modes in the two baths. It is subsequently found that the approach does not work so well for the case of $s<0.5$ due to the large truncation error. In comparison, our approach with SOPB can be applied to all cases, especially for the cases with $s<0.5$ and $\alpha_z\neq\alpha_x$.

It was stated that in the numerical calculations the U($1$) symmetry will always be broken spontaneously \cite{Guo3}. Indeed, our DMRG results do not show simultaneous magnetization disappearance in both $Z$ and $X$ orientations, i.e., $\langle\sigma^z\rangle=\langle\sigma^x\rangle=0$. Whereas, the results of $\zeta$ show that the symmetry is still held. This shows that the involvement of the bosonic component in the order parameter is essential. To confirm this statement, we adopt the variational approach to investigate the system with two single-mode bosonic baths (i.e., $M=1$ in Eq.~(\ref{vmwave})), which could be dealt with in a numerically exact manner. We will show that in the exact case, the symmetry is spontaneously broken in the strong coupling case with $\lambda \gg \omega$.

The ground state wave function of the TBSBM is also investigated to reveal the nature of the localized and delocalized phases.
The phonon population $P(x,z)$ is introduced to describe the ground state of the two-bath model.
Assuming that the wave function of the ground state can be written as
\begin{equation}
|\Psi_{g}\rangle =  | + \rangle |\psi_{+}\rangle_{\rm ph} +  |-\rangle |\psi_{-}\rangle_{\rm ph},
\label{gs}
\end{equation}
where $|\psi_{+}\rangle_{\rm ph}$ and $|\psi_{-}\rangle_{\rm ph}$ are the phonon parts of the wave function corresponding to the spin up and down states, respectively, which can be expanded in a series of Fock states or coherent states. Thus the phonon population $P(x,z)$ is defined as
\begin{eqnarray}
P(x,z) & = & \langle \Psi_g(x,z)|\Psi_g(x,z)\rangle_{\rm spin} \nonumber  \\
& = &   |\psi_{+}(x,z)|^2 + |\psi_{-}(x,z)|^2
\label{density}
\end{eqnarray}
where $\langle \cdots \rangle_{\rm spin}$ represents the trace of the spin freedom of the wave function, $x$ and $z$ correspond to the off-diagonal and diagonal coupling baths, respectively, and $\psi_{\pm}(x,z)=\langle\vec{r}|\psi_{\pm}\rangle_{\rm ph}$ is the phonon part of the wave function in the two-dimensional coordinate representation $\vec{r}=(x, z)$.

According to the definitions of the phonon population $P(x,z)$ in Eq.~(\ref{density}) and the variational ansatz in Eq.~(\ref{vmwave}),  $P(x,z)$ can be calculated as
\begin{equation}
P(x,z)=\sum_{n=1}^{N}\frac{\left[A_nF_{n}(x,z)\right]^2+\left[B_nG_{n}(x,z)\right]^2}{D},
\label{vm_density}
\end{equation}
where $D=\langle\Psi_g|\Psi_g\rangle$ is the norm of the wave function, $A_n$ and $B_n$ denote the weight coefficients of the $n$-th coherent superposition state coupled to the spin up and down states, respectively, and $F_n(x,z)=\langle\vec{r}|\psi_{+}\rangle_{\rm ph}=f_{n,x}(x)f_{n,z}(z)$ and $G_n(x,z)=\langle\vec{r}|\psi_{-}\rangle_{\rm ph}=g_{n,x}(x)g_{n,z}(z)$ represent the phonon part of the wave function $|\psi_{\pm}\rangle_{\rm ph}$ in the coordinate representation $\vec{r}=(x,z)$. The function $f_{n,x}(x)$ denoting the phonon state in the off-diagonal coupling bath then can be obtained,
\begin{eqnarray}
f_{n,x}(x) & = & \prod_l \langle x | f_{n,l} \rangle   \\
& = & \prod_l \left( \frac{\omega_l}{\pi}\right)^{1/4} e^{-ix_lp_l/2}e^{ip_lx}e^{-\omega(x-x_l)^2/2},  \nonumber
\end{eqnarray}
where $x_l$ and $p_l$ are defined as
\begin{eqnarray}
  p_l &  = & -i \sqrt{\frac{\omega_l}{2}}\left(f_{n,l} - f_{n,l}^{*}\right), \\
  x_l & = & \frac{1}{\sqrt{2\omega_l}}\left(f_{n,l} + f_{n,l}^{*}\right).
\end{eqnarray}
In the same way, the functions $f_{n,z}(z), g_{n,x}(x)$ and $g_{n,z}(z)$ can also be calculated from the displacement coefficients $f_{n,l}$ and $g_{n,l}$.
Without loss of generality, we set the frequency $\omega_l$ to unity such that $X$ and $Z$ are now dimensionless.
In the single-mode case with the number of effective bath modes $M=1$, the function $F_n(x,z)=f_{n,x}(x)f_{n,z}(z)=\langle z | f_{n,1} \rangle \langle x | f_{n,2} \rangle$ is
simplified, where the subscript $1$ and $2$ correspond to the diagonal and off-diagonal coupling baths, respectively.

\begin{figure}
\includegraphics[angle=0,scale=1.6]{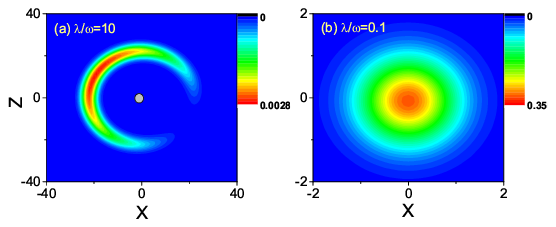}
\caption{ The wave function of the ground state for the single-mode case (i.e., M=1 in Eq.~(\ref{vmwave})) is plotted in (a) for the strong coupling
regime with $\lambda/\omega=10$ and in (b) for the weak coupling regime with $\lambda/\omega=0.1$. The $X$-coordinate and $Z$-coordinate
correspond to the off-diagonal and diagonal coupling baths, respectively, and the colour represents the distribution
of the phonon population $P(x, z)$.
}
\label{variation_8}
\end{figure}

As shown in Fig.~\ref{variation_8} (a), the distribution of the phonon probability density $P(x,z)$ is located at a corner
of the $X$-$Z$ plane for the strong coupling regime with $\lambda/\omega=10$, corresponding to the localized phase. Quite different from that in (a),
the probability density distribution of the phonon $P(x,z)$ in the critical phase concentrates at the origin, as shown in Fig.~\ref{variation_8} (b) for the weak coupling regime with $\lambda/\omega=0.1$. It shows that the distance between the center of $P(x,z)$ and the origin is an important parameter to watch as one tells the localize phase from the critical one.